\documentstyle[preprint,aps]{revtex}
\tighten
\draft
\title{From number theory to statistical mechanics: \\
    Bose--Einstein condensation in isolated traps}
\author{Siegfried Grossmann and Martin Holthaus}
\address{Fachbereich Physik der Philipps-Universit\"at,
    Renthof 6, D-35032 Marburg, Germany}
\date{August 6, 1997}
\narrowtext
\begin{document}

\maketitle

\begin{abstract}
We question the validity of the grand canonical ensemble for the description
of Bose--Einstein condensation of small ideal Bose gas samples in isolated
harmonic traps. While the ground state fraction and the specific heat capacity
can be well approximated with the help of the conventional grand canonical
arguments, the calculation of the fluctuation of the number of particles
contained in the condensate requires a microcanonical approach. Resorting to
the theory of restricted partitions of integer numbers, we present analytical
and numerical results for such fluctuations in one-- and three-dimensional
traps, and show that their magnitude is essentially independent of the total
particle number.
\end{abstract}

\section{Bose--Einstein condensation in isolated traps: Can we rely on the
grand canonical ensemble?}

After the breakthrough in the preparation and detection of Bose--Einstein
condensates of rubidium~\cite{AndersonEtAl95}, sodium~\cite{DavisEtAl95},
and lithium~\cite{BradleyEtAl97}, the second generation of experiments on
ultracold, magnetically trapped Bose gases now starts to probe condensate
properties such as the temperature dependence of the condensate
fraction~\cite{MewesEtAl96a,EnsherEtAl96} or collective excitations of the
condensate~\cite{JinEtAl96,MewesEtAl96b,JinEtAl97}. A matter-of-principle
experiment has shown the feasibility of an output coupler for condensed
trapped atoms~\cite{MewesEtAl97}, two different, overlapping condensates have
been produced in a single trap~\cite{MyattEtAl97}, and interference fringes
between two freely expanding condensates have been
demonstrated~\cite{AndrewsEtAl97a}, thus proving first-order coherence of
``mesoscopic'' matter waves. Higher-order coherence has been inferred from
three-body recombination rates of condensed atoms~\cite{BurtEtAl97}, and
even the propagation of sound in a Bose condensate could be directly
observed~\cite{AndrewsEtAl97b}. 

These developments naturally force the theorist to investigate a conceptually
important question: When dealing with Bose--Einstein condensates consisting of
about $10^3$ to $10^6$ particles in a magnetic trap, to which extent can the
different thermodynamical ensembles be considered as equivalent? A Bose gas
cooled below the condensation point and kept in a trap is neither in thermal
contact with a heat bath, nor exchanging particles with some particle
reservoir. Hence, the only ensemble that is appropriate for describing this
situation is the microcanonical one. And yet, the theoretical discussion of
trapped condensates so far mainly relies on the grand canonical ensemble. 

Within the grand canonical ensemble, the description of a trapped, ideal Bose
gas consisting of $N$ particles is comparatively simple. We consider an
isotropic harmonic trap with oscillator frequency $\omega$, so that the
degree of degeneracy of the $j$-th single-particle state is 
\begin{equation}
g_j = \frac{1}{2}(j+1)(j+2) \; .
\label{DEG}
\end{equation}  
Hence, the difference $\widetilde{\mu}$ of the chemical potential and
the single-particle ground state energy is determined by the equation
\begin{equation}
N = \sum_{j=0}^{\infty} \frac{j^2/2 + 3j/2 + 1}
    {\exp[(j\hbar\omega - \widetilde{\mu})/k_BT] - 1}   \; ,
\end{equation}
where $k_B$ denotes the Boltzmann constant, and $T$ is the temperature.
This sum can be evaluated approximately by introducing the density of states
\begin{equation}
\rho(E) \approx \frac{1}{2} \frac{E^2}{(\hbar\omega)^3}
    + \gamma \frac{E}{(\hbar\omega)^2}
\label{RHO}
\end{equation} 
with $\gamma = 3/2$, which follows immediately from eq.~(\ref{DEG}).
Converting the sum into an integral and applying standard arguments, one then
finds the condensation temperature~\cite{GH95a,GH95b,KetterleDruten96}
\begin{equation}
T_C^{(3)} \approx T_0^{(3)}
    \left[1 - \frac{\gamma \, \zeta(2)}{3 \, \zeta(3)^{2/3}}
    \cdot \frac{1}{N^{1/3}}\right] \; ,
\label{TCO}
\end{equation}
where
\begin{equation}
T_0^{(3)} = \frac{\hbar\omega}{k_B} \left( \frac{N}{\zeta(3)} \right)^{1/3}	
\end{equation}
denotes the condensation temperature pertaining to the
large-$N$-limit~\cite{deGrootEtAl50,BagnatoEtAl87}; $\zeta(z)$ is the
Riemann zeta function. The lowering of $T_C^{(3)}$ with respect to
$T_0^{(3)}$ is of order $N^{-1/3}$ and results from the enhancement of the
density of states~(\ref{RHO}) above the leading ``volume''-term: since there
are more states available, the need to condense arises only at lower
temperatures. It should be noted that, strictly speaking, there is no
well-defined condensation temperature for a gas consisting of a finite
number of particles. However, Fig.~1 clearly demonstrates that the onset
of condensation in a three-dimensional harmonic trap becomes quite sharp
already for particle numbers of the order of~$10^5$, and eq.~(\ref{TCO})
describes this onset very well.

In contrast to the textbook case of the free Bose gas, the heat capacity
of the harmonically trapped Bose gas exhibits a steep drop at the condensation
point. As shown in Fig.~2, this drop becomes a discontinuous jump by about
$6.6~k_B$ per particle in the large-$N$-limit.

Remaining within the scope of the grand canonical ensemble,
this analysis can be made formally more precise, and generalized
to moderately anisotropic harmonic
traps~\cite{KirstenToms96a,KirstenToms96b,HaugerudEtAl97,HaugsetEtAl97},
which changes the value of $\gamma$. But can one rely on the grand canonical
ensemble? Ziff, Uhlenbeck, and Kac have made a case that {\em the grand
canonical ensemble does not represent any physical situation for the condensed
ideal Bose gas\/}, and advocate that {\em its anomalous predictions should be
ignored\/}~\cite{ZiffEtAl77}. These ``anomalous predictions'' are related to
what we call the ``grand canonical fluctuation catastrophe'': within the grand
canonical ensemble, the r.m.s.-fluctuations $\delta N_0$ of the ground state
occupation number $N_0$ are given by~\cite{LandauLifshitz59,Pathria85}
\begin{equation}
\left(\delta N_0\right)^2 = N_0\left(N_0 + 1 \right)	\; , 
\end{equation}    
implying that $\delta N_0$ is of the order of the total particle number $N$
below the condensation point. However, if we consider an isolated Bose gas
in a trap, all particles occupy the ground state at zero temperature, so that
the true fluctuations of $N_0$ vanish. The grand canonical prediction for
$\delta N_0$ thus differs drastically from the microcanonical one.

\section{Microcanonical approach to fluctuations of the ground state
    occupation}

Since the magnitude of the fluctuations $\delta N_0$ is related to the
coherence properties of the condensate, it is of substantial interest to
compute the true, i.e., microcanonical fluctuations for a trapped ideal Bose
gas. We will study an isotropic harmonic trapping potential with oscillator
frequency $\omega$ in $d$ dimensions, and denote by $n$ the number of
excitation quanta for some preassigned value of the excitation energy $E$:
\begin{equation}
n = \frac{E}{\hbar\omega}   \; .
\end{equation}
The prime task now is to determine the number $\Omega^{(d)}(n|N)$ of
microstates. Since there are generally many microstates where only a
part of the $N$ particles carries all $n$ excitation quanta, leaving
the other particles in the ground state, $\Omega^{(d)}(n|N)$ equals
the number of possibilities for distributing the $n$ quanta over
{\em at most\/} $N$ Bose particles. Then the difference
$\Omega^{(d)}(n|N_{\rm ex}) - \Omega^{(d)}(n|N_{\rm ex}-1)$ is the
number of possibilities for distributing $n$ quanta over {\em exactly\/}
$N_{\rm ex}$ particles, with $N_{\rm ex}$ ranging from $1$ to $N$, so that   
\begin{equation}
p_{\rm ex}^{(d)}(N_{\rm ex}|n) =
\frac{\Omega^{(d)}(n|N_{\rm ex}) - \Omega^{(d)}(n|N_{\rm ex}-1)}
{\Omega^{(d)}(n|N)} \; , \qquad N_{\rm ex} = 1,2,\ldots,N \; ,
\label{DIS}
\end{equation}
is the probability for finding $N_{\rm ex}$ out of $N$ particles excited
when the total excitation energy is $n \cdot \hbar\omega$. Since the
remaining $N - N_{\rm ex}$ particles occupy the ground state, the first
moment $\langle N_{\rm ex} \rangle$ of the distribution~(\ref{DIS}) yields
die microcanonical expectation value of the ground state occupation number
according to
\begin{equation}
\langle N_0 \rangle = N - \langle N_{\rm ex} \rangle \; ;      
\end{equation}
the corresponding fluctuations follow from
\begin{equation}
\left(\delta N_0\right)^2 =
    \langle N_{\rm ex}^2 \rangle - \langle N_{\rm ex} \rangle^2	\; . 
\label{FLU}
\end{equation}

\section{Condensate fluctuations in a one-dimensional oscillator potential}

The case $d=1$ had been considered already in 1949 by
Temperley~\cite{Temperley49}, and shortly there\-after by Nanda~\cite{Nanda53}.
It is of particular interest, since it can be treated in detail analytically
with the tools furnished by the theory of the partion of integers: the number
$\Omega^{(1)}(n|N)$ of microstates corresponds to the number of partitions
of the integer $n$ into at most $N$ summands; the commutativity of the
summands reflects the indistinguishability of the Bosons. We now have to
distinguish two cases: if $n \le N$, then the fact that the number of
particles is finite has no consequences for the enumeration of microstates,
and we are dealing with {\em unrestricted\/} partitions of $n$. If $n > N$,
the partitions of $n$ are {\em restricted\/} by the requirement that the
number of summands, corresponding to the number of excited particles, does
not exceed $N$.

Let us first consider the case $n \le N$. We follow the usual convention
and denote the number of unrestricted partitions of $n$ as $p(n)$. 
Introducing the dimensionless inverse temperature $\xi = \hbar\omega/(k_B T)$,
it is an elementary exercise to show that 
\begin{equation}
\sum_{n=0}^{\infty} p(n) \, e^{-n\xi}
    = \prod_{j=1}^{\infty}\frac{1}{1 - \exp(-j\xi)}
    \equiv Z^{(1)}_{\infty}(\xi) \; ,
\label{GEN}
\end{equation}
which means that the generating function for $p(n)$ corresponds, physically
speaking, to the canonical partition function of a fictituous system of
infinitely many distinguishable harmonic oscillators with frequencies that
are integer multiples of $\omega$. Given this, it is natural to introduce
the new variable $x = \exp(-\xi)$ and to extract the numbers $p(n)$ by
inverting eq.~(\ref{GEN}) with the help of the saddle point approximation,
after expressing $\xi$ in terms of $n$. But we have to be careful: for
temperatures that may physically be considered as ``low'', but are still high
compared to $\hbar\omega/k_B$, the variable $\exp(-\xi)$ is close to unity,
so that we need to know the behaviour of $Z^{(1)}_{\infty}(\xi)$ in the
vicinity of a singularity. However, there exists the remarkable
identity (see eq.~(1.42) in ref.~\cite{HardyRamanujan18})
\begin{equation}
Z^{(1)}_{\infty}(\xi) = \sqrt{\frac{\xi}{2\pi}} \,
    \exp\!\left(\frac{\zeta(2)}{\xi} - \frac{\xi}{24}\right) \,
    Z^{(1)}_{\infty}(4\pi^2/\xi)
\end{equation}
that links the low-temperature behaviour of our fictituous oscillator
system to its high-temperature dynamics. This allows us to derive
the low-temperature approximation
\begin{equation}
\ln Z^{(1)}_{\infty}(\xi) \approx \frac{\zeta(2)}{\xi}
    + \frac{1}{2}\ln\xi - \frac{1}{2}\ln 2\pi	\; ,
\end{equation}
which, in turn, yields the desired energy-temperature relation:
\begin{equation}
n \; = \; -\frac{\partial}{\partial \xi} \ln Z^{(1)}_{\infty}(\xi)
    \; \approx \; \frac{\zeta(2)}{\xi^2}  \; .
\end{equation}
Now we can compute the canonical entropy
\begin{equation}
S(\xi)/k_B = n\xi + \ln Z^{(1)}_{\infty}(\xi)
\end{equation}
and apply ``Bethe's theorem''~\cite{Bethe37} to get
\begin{equation}
p(n) \; \sim \; \frac{\exp\left[S(\xi(n))/k_B\right]}
    {\left(-2\pi\frac{\partial n}{\partial \xi}\right)^{1/2}}
    \; = \; \frac{1}{4\sqrt{3}n}\exp\!\left(\pi\sqrt{\frac{2}{3}n}\right) \; .
\end{equation}
The right hand side is nothing but the celebrated Hardy--Ramanujan formula
for the number of unrestricted partitions~\cite{HardyRamanujan18}, so we have
just sketched a physicist's solution to a number-theoretical problem. There is
one detail that deserves particular attention: Bethe's theorem amounts to the
inversion of the canonical partition sum~(\ref{GEN}) within the saddle
point approximation, so that the dominant contributions to this sum are
properly taken into account within the Gaussian approximation. As a result,
the microcanonical entropy $\ln p(n)$ differs from the canonical entropy
$S(\xi(n))/k_B$ by the saddle point correction
$-\ln(-2\pi\partial n/\partial\xi)/2$. Such differences between
thermodynamical quantities pertaining to different ensembles are
characteristic for small systems. Expressed the other way round, the usually
assumed equality of thermodynamical quantites in different ensembles holds
to the extent that such saddle point corrections can be neglected.  
 
Now we can turn to the case $n > N$, where the number of quanta
exceeds the number of particles, and have to determine the number
$\Omega^{(1)}(n|N) \equiv p_N(n)$ of restricted partitions of $n$.
In principle, one can proceed as in the previous case, since there is
the identity~\cite{AuluckKothari46}
\begin{equation}
\sum_{n=0}^{\infty} p_N(n) \, e^{-n\xi}
    = \prod_{j=1}^{N}\frac{1}{1 - \exp(-j\xi)}
    \equiv Z^{(1)}_{N}(\xi) \; :
\label{RES}
\end{equation}
The generating function for $p_N(n)$, that is, the canonical partition
function for $N$ ideal Bosons trapped by a one-dimensional harmonic
potential, equals the canonical partition function of a system of
$N$~harmonic oscillators with frequencies $\omega$, $2\omega$, $\ldots \,$,
$N\omega$. The corresponding asymptotic formula for $p_N(n)$, which
turns out to be rather intricate, has been given by Auluck and
Kothari~\cite{AuluckKothari46}. However, a beautiful theorem due to
Erd\"os and Lehner~\cite{ErdosLehner41} helps to simplify the analysis:	
if, for some given $n$ and $x$, the number $N$ obeys
\begin{equation}
N = \frac{\sqrt{n}\ln n}{C} + x\sqrt{n}
    \qquad {\mbox{with}} \qquad C = \pi\sqrt{\frac{2}{3}}	\; , 
\end{equation}
then
\begin{equation}
\lim_{n\to\infty} \frac{p_N(n)}{p(n)} \; = \;
    \exp\!\left(-\frac{2}{C}e^{-\frac{1}{2}Cx}\right) \; .
\end{equation}
Hence, for given large values of $n$ and $N$ we {\em define\/} $x$ according to
\begin{equation}
x = \frac{N}{\sqrt{n}} - \frac{\ln n}{C} \; ,
\end{equation}
and obtain the approximation
\begin{equation}
p_N(n) \approx p(n) \, \exp\!\left(-\frac{2}{C}e^{-\frac{1}{2}Cx}\right) \; .
\label{ELF}
\end{equation}
In order to check this approximation, we first determine the numbers
$p_N(n) = \Omega^{(1)}(n|N)$ by means of the saddle point inversion of
eq.~(\ref{RES}), evaluate the distributions~(\ref{DIS}), and compute the
relative fluctuations $\delta N_0/N$ according to eq.~(\ref{FLU}). The
results for $N = 10^3$, $10^4$, and $10^5$ are shown as full lines in Fig.~3.
The corresponding data obtained with the help of the Erd\"os--Lehner
approximation~(\ref{ELF}) are indicated by the dashed lines. Evidently,
the approximation is quite good already for $N = 10^4$.

The merit of the approximation~(\ref{ELF}) lies in the fact that it
allows us to derive an analytical expression for the microcanonical
low-temperature ground state fluctuations of the ideal Bose gas trapped
by a one-dimensional harmonic potential~\cite{GH96}:
\begin{equation}   
\delta N_0 \approx \frac{\pi}{\sqrt 6} \frac{k_B T}{\hbar\omega}
    \quad {\mbox{for}} \quad
    T \ll T_0^{(1)} \equiv \frac{\hbar\omega}{k_B} \frac{N}{\ln N}   \; .  
\label{GSF}
\end{equation}
As already seen in Fig.~3, the low-temperature fluctuations vanish
linearly with temperature. Most notably, they do {\em not\/} depend on
the particle number $N$. This is immediately obvious for temperatures
below the ``restriction temperature'' $T_R^{(1)}$ that is defined by the
condition $n(T_R^{(1)}) = N$, since below $T_R^{(1)}$ the number of
microstates becomes independent of $N$, so that ``the condensate has no
chance to know how many particles it consists of''. However, for $d = 1$
this restriction temperature is merely of the order of $N^{1/2}$, since
$k_B T_R^{(1)} \approx \hbar\omega(N/\zeta(2))^{1/2}$,
wheras the $N$-independence of $\delta N_0$ actually persists almost up to
$k_B T_0^{(1)} = \hbar\omega (N/\ln N)$, where the occupation of the ground
state becomes significant~\cite{GH96}. This $N$-independence of $\delta N_0$
appears to be characteristic for isolated condensates in general, as we
shall indicate below.

\section{Condensate fluctuations in a three-dimensional trap}

The analysis for the case $d = 3$ proceeds in close analogy to the
case $d = 1$. The restriction temperature, determined from
$n(T_R^{(3)}) = N$, now reads
\begin{equation}
T_R^{(3)} \approx 0.8 \, N^{-1/12} \, T_0^{(3)}	\; ,
\end{equation}
which means that $T_R^{(3)}$ is about one fourth of the condensation
temperature for a gas consisting of $10^6$ particles. Below $T_R^{(3)}$
the number of microstates, and hence all thermodynamical properties of
the trapped gas, are independent of the particle number $N$. Accordingly,
we write $\Omega^{(3)}(n|N) = \Omega^{(3)}(n)$ for $n$ corresponding to
temperatures~$T$ less than $T_R^{(3)}$, so that $\Omega^{(3)}(n)$ denotes
the three-dimensional analogue of $p(n)$. The generating function for
$\Omega^{(3)}(n)$ again equals the canonical partition function of a system
of infinitely many, distinguishable harmonic oscillators~\cite{GH97}:
\begin{equation}
\sum_{n=0}^{\infty} \Omega^{(3)}(n) \, e^{-n\xi}
    = \prod_{j=1}^{\infty}\frac{1}{[1 - \exp(-j\xi)]^{g_j}}
    \equiv Z^{(3)}_{\infty}(\xi) \; .
\label{OSC}
\end{equation}
The saddle point inversion of this generating function is straightforward, but
a bit tedious. The result has been given by Nanda in 1951~\cite{Nanda51}:
\begin{equation}
\Omega^{(3)}(n) =
\frac{\widetilde{n}^{-25/32}}{4\pi\left(3\zeta(4)\right)^{1/2}}
        \exp\!\left(4\zeta(4)\widetilde{n}^{3/4}
      + \frac{3}{2}\zeta(3)\widetilde{n}^{1/2}
      + \left[\zeta(2) - \frac{3}{8}\frac{\zeta(3)^2}{\zeta(4)}\right]
	\widetilde{n}^{1/4} + B \right)	  
\label{NAN}
\end{equation} 
with
\[
\widetilde{n} = \frac{n}{3\zeta(4)}
\]
and
\[
B = \frac{\zeta(3)^3}{8\zeta(4)^2} - \frac{\zeta(2)\zeta(3)}{4\zeta(4)}
        + \frac{3}{2}\zeta'(-1) + \frac{1}{2}\zeta'(-2)	\;.
\]
This is one of the rare examples of a truly asymptotic formula for the
number of microstates of a non-trivial Bose system.

For temperatures above $T_R^{(3)}$ the finiteness of the particle number
restricts the number of microstates, and we have to compute
$\Omega^{(3)}(n|N)$. It turns out that the logarithm of $\Omega^{(3)}(n)$
provides a quite good approximation to $\ln\!\left(\Omega^{(3)}(n|N)\right)$
even up to the condensation temperature~\cite{GH97}, so that the entropy
of the fictituous Boltzmannian oscillator system described by the
partition function~(\ref{OSC}) approximates the entropy of the trapped
Bose gas for all temperatures below the onset of condensation. This means
that the distributions $p_{\rm ex}^{(3)}(N_{\rm ex}|n)$ introduced in
eq.~(\ref{DIS}) must be well peaked; those microstates where the $n > N$
quanta are actually ``spread out'' over all $N$ particles carry only a minor
statistical weight below $T_C^{(3)}$. But still, this finding is of no help
for computing the fluctuations $\delta N_0$, since this computation requires,
according to eq.~(\ref{DIS}), to consider auxiliary systems consisting of
$N_{\rm ex} \le N$ particles and to determine the numbers
$\Omega^{(3)}(n|N_{\rm ex})$. The problem now is that an immediate
analogue to the generating function~(\ref{RES}) does not exist, that is,
the canonical $N$-particle partition function $Z^{(3)}_{N}(\xi)$ is not
known in closed form. However, there exists the recursion
formula~\cite{BorrmannFranke93,Eckhardt97,WilkensWeiss97}
\begin{equation}
Z_N^{(d)}(\xi) = \frac{1}{N}\sum_{k=1}^{N}
    Z_1^{(d)}(k\xi) Z^{(d)}_{N-k}(\xi)
\label{REC}
\end{equation}
that can numerically be evaluated and inverted within the saddle point
approximation, so that we can determine the distributions
$p_{\rm ex}^{(3)}(N_{\rm ex}|n)$ at least numerically with high
precision~\cite{GH97}. The results for $N = 1000$ are shown in
Fig.~4: as anticipated, the distributions are well peaked, and moreover
remarkably close to Gaussians.  

From the expectation values of these distributions we obtain the
microcanonical expectation values of the ground state occupation number,
depicted as the full line in Fig.~5. The agreement with the corresponding
grand canonical data (dashed) is stunning, but in view of the preceding
Fig.~4 not unexpected: the predictions of the different ensembles will
be the same as long as the most probable values of $N_{\rm ex}$ coincide
with the expectation values, which is the case as long as the
distributions $p_{\rm ex}^{(3)}(N_{\rm ex}|n)$ stay symmetrical. This,
in turn, is guaranteed as long as the support of these distributions
stays away from the border $N_{\rm ex}/N = 1$. Hence, there is a
slight difference between the microcanonical and the grand canonical
ground state fraction only in the immediate vicinity of the condensation
point. For $d = 1$ this difference is much more pronounced, since
$p_{\rm ex}^{(1)}(N_{\rm ex}|n)$ is considerably less well peaked~\cite{GH96}
than its counterpart for $d = 3$.

Fig.~6 shows the microcanonical fluctuations of the ground state
occupation number for $d = 3$ and $N = 200$, $500$, and $1000$. Remarkably,
the low-temperature fluctuations for these three systems agree perfectly.
More generally, if we compare two trapped Bose gases with different $N$
under otherwise identical conditions, then their fluctuations $\delta N_0$
are nearly identical for temperatures below the lower of their condensation
temperatures. This finding generalizes the formula~(\ref{GSF}). Again, the
$N$-independence of $\delta N_0$ is immediately obvious for temperatures
lower than the restriction temperature $T_R^{(3)}$, but it actually persists
almost up to the condensation temperature, as a result of the well-peakedness
of the distributions $p_{\rm ex}^{(3)}(N_{\rm ex}|n)$.

\section{Concluding remarks}

We now come back to our initial question: Can we rely on the grand
canonical ensemble for Bose--Einstein condensation in isolated traps?
As long as we are only interested in the usual thermodynamical quantities,
the answer is yes. Fig.~5 has shown that the grand canonical prediction
for the ground state occupation number in a three-dimensional trap is almost
indistinguishable from the microcanonical one even if the particle number
$N$ is as low as $1000$. But in view of the fact that one-dimensional traps
are not out of reach~\cite{KetterleDruten96,DrutenKetterle97}, it should be
kept in mind that these differences are larger for $d = 1$~\cite{GH96}.

Another comparison between different ensembles is presented in Fig.~7,
which depicts the grand canonical (long dashes), canonical (short dashes),
and microcanonical (full line) specific heat capacity for $d = 3$ and
$N = 1000$. Again, noticeable differences can be found only close to
the onset of condensation.

But when it comes to quantum statistical properties, the grand canonical
ensemble is unsound~\cite{ZiffEtAl77}. Adopting the microcanonical spirit
and starting from the enumeration of microstates, we have computed the
microcanonical low-temperature fluctuations $\delta N_0$ for harmonically
trapped Bose gases. While the case $d = 1$ could be treated fully
analytically, leading to eq.~(\ref{GSF}), the case $d = 3$ had to rely
on the numerical evaluation of the recursion relation~(\ref{REC}). If one
could derive an analytical approximation to the numbers $\Omega^{(3)}(n|N)$
for $n > N$, with a quality similar to that of the Erd\"os--Lehner
approximation~(\ref{ELF}) to $\Omega^{(1)}(n|N) \equiv p_N(n)$, then the case
$d = 3$ could be brought into the same textbook status that the case $d = 1$
already has by now.

It is also worthwhile to point out that very recently Navez {\em et al.\/}
have proposed a new statistical ensemble, within which the ground state
particles act as a reservoir, and exchange of particles with the
excited-states subsystem without exchange of energy is
possible~\cite{NavezEtAl97}. The predictions for the fluctuations $\delta N_0$
obtained with the help of this so-called ``Maxwell's Demon ensemble'' agree
very favourably with our strictly microcanonical results, so that this
ensemble might constitute a valuable tool for further studies. 

In closing, we reiterate the most important result of the present analysis:
while the grand canonical ensemble predicts, for ideal Bose gases at low
temperatures, ground state number fluctuations $\delta N_0$ of the order $N$,
and while usual thermodynamical fluctuations scale as~$\sqrt{N}$, the 
low-temperature fluctuations $\delta N_0$ for trapped, isolated ideal Bose
condensates are independent of the total particle number $N$.

\begin{figure}
\caption[FIG.~1] {Grand canonical ground state fraction $N_0/N$ for a
    three-dimensional isotropic harmonic trap, and gases consisting of $10^3$,
    $10^5$, and $10^7$ ideal Bose particles. The larger the particle number,
    the sharper the onset of condensation, and the smaller the shift of the
    condensation temperature~(\ref{TCO}) with respect to $T_0 = T_0^{(3)}$.
    These data result from exact numerical calculations that do not invoke
    the continuum approximation.} 
\end{figure}

\begin{figure}
\caption[FIG.~2] {Grand canonical specific heat capacities for harmonically
    trapped ideal Bose gases ($d=3$). The particle numbers are $N = 10^3$,
    $10^4$, $\ldots \,$, $10^7$. The higher the particle number, the
    steeper the drop at the onset of condensation.}
\end{figure}

\begin{figure}
\caption[FIG.~3] {Microcanonical fluctuations of the ground state
    occupation number for ideal Bose gases in a one-dimensional harmonic
    trap, for $N = 10^3$, $10^4$, and $10^5$ (top to bottom).
    Full lines: data obtained with the help of the saddle point inversion
    of eq.~(\ref{RES}).
    Dashed: data resulting from the Erd\"os--Lehner formula~(\ref{ELF}).
    $T_0 = T_0^{(1)} \equiv \hbar\omega N/(k_B\ln N)$ marks the temperature
    below which the ground state occupation becomes significant~\cite{GH96}.
    The seeming $N$-dependence of the low-temperature fluctuations stems
    from the fact that they have been plotted versus the reduced temperature
    $T/T_0^{(1)}$, since $T_0^{(1)}$ depends on $N$.}
\end{figure}

\begin{figure}
\caption[FIG.~4] {Microcanonical probability distributions
    $p_{\rm ex}^{(3)}(N_{\rm ex}|n)$ for finding $N_{\rm ex}$ out of
    $N = 1000$ isotropically trapped Bose particles excited when the total
    energy is $n \cdot \hbar\omega$. The temperatures $T/T_0^{(3)}$
    corresponding to the Gaussian-like distributions range from $0.3$ to
    $0.9$ (left to right, in steps of $0.1$); the temperature for the
    rightmost, monotonous distribution is $T = 0.95 \, T_0^{(3)}$,
    which is higher than $T^{(3)}_C \approx 0.93 \, T_0^{(3)}$,
    see eq.~(\ref{TCO}).}
\end{figure}

\begin{figure}
\caption[FIG.~5] {Microcanonical ground state fraction (full line)
    for $N = 1000$ isotropically trapped Bose particles ($d=3$) versus
    reduced temperature $T/T_0^{(3)}$, compared to the corresponding
    grand canonical data (dashed).}
\end{figure}

\begin{figure}
\caption[FIG.~6] {Microcanonical fluctuations $\delta N_0$ versus
    temperature, for $d = 3$ and $N = 200$, $500$, and $1000$. The
    fluctuations are maximal close to the respective condensation points.
    Below the lowest of the condensation points, the fluctuations of the
    three systems agree perfectly, thus revealing the $N$-independence
    of $\delta N_0$.}   
\end{figure}

\begin{figure}
\caption[FIG.~7] {Comparison of the grand canonical (long dashes),
    canonical (short dashes), and microcanonical (full line) specific
    heat capacity for $d = 3$ and $N = 1000$.}
\end{figure}
   
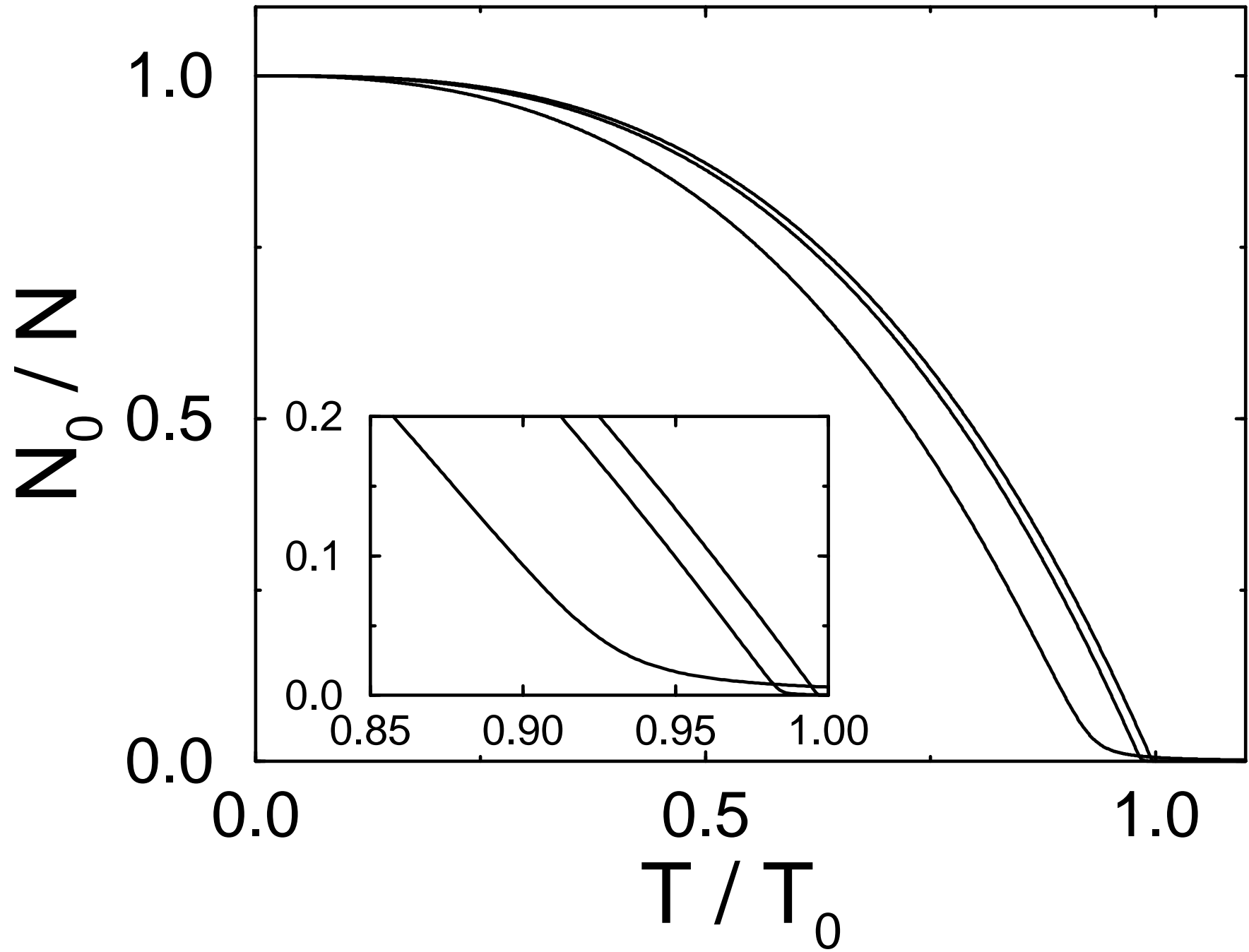
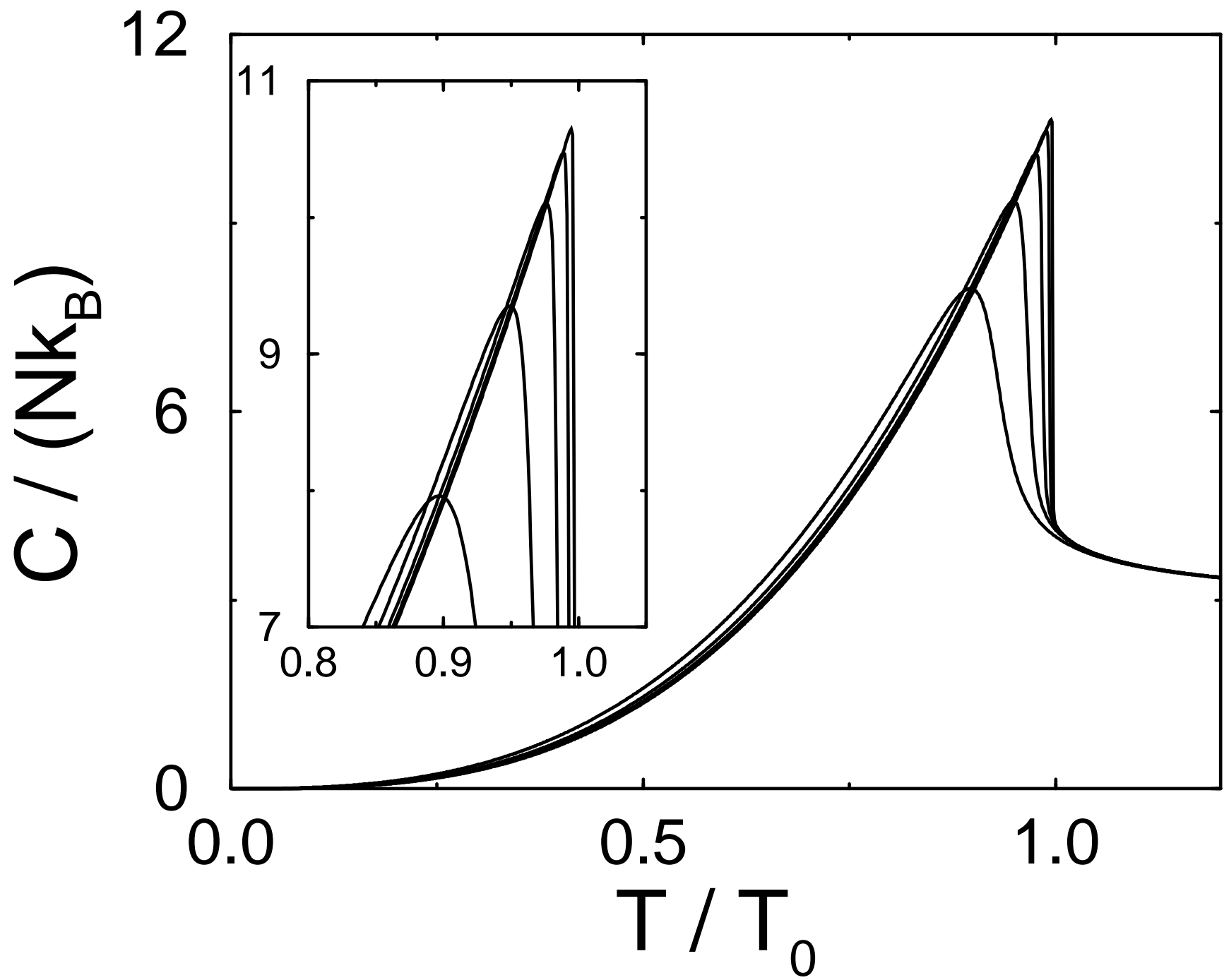
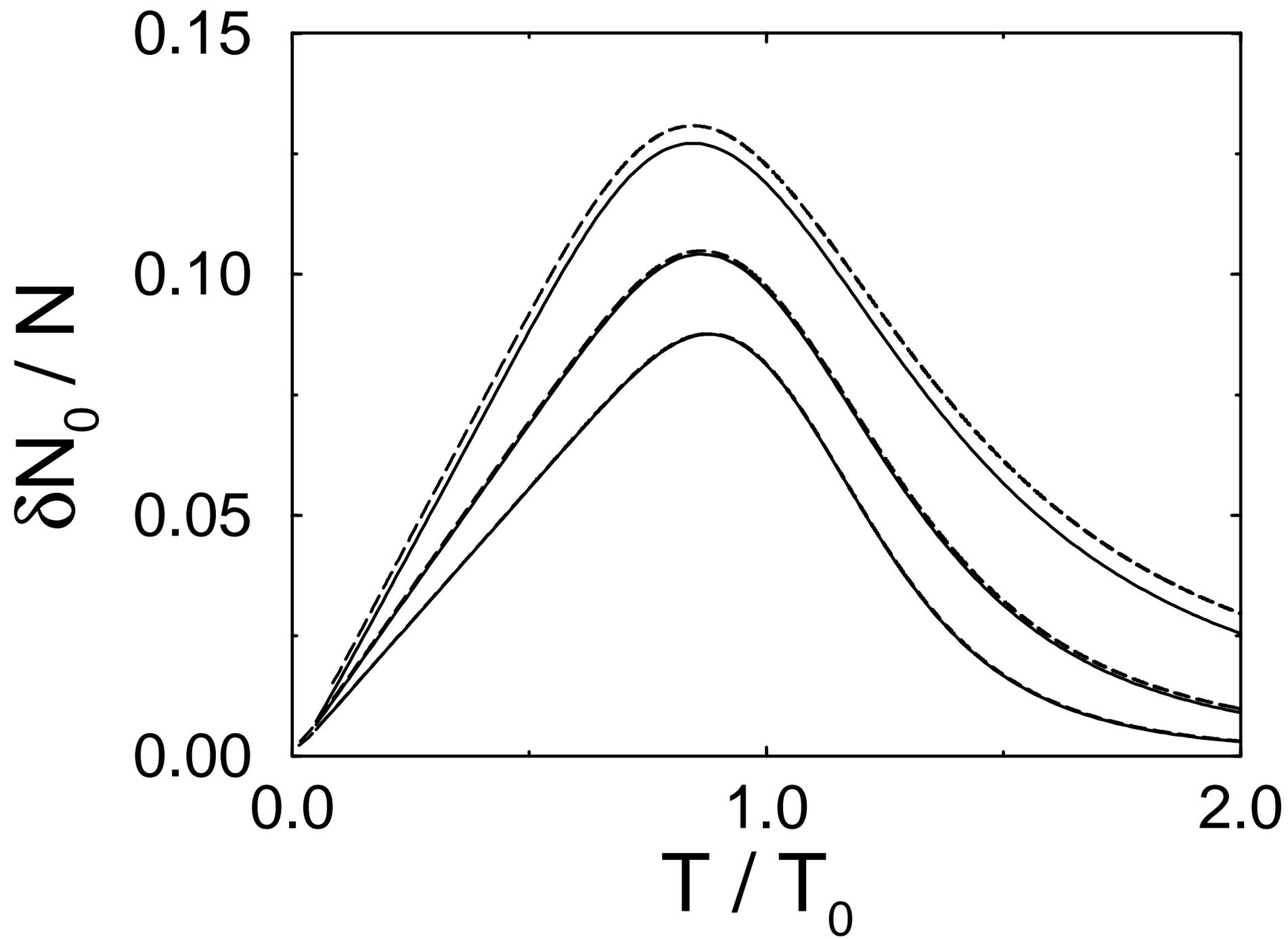
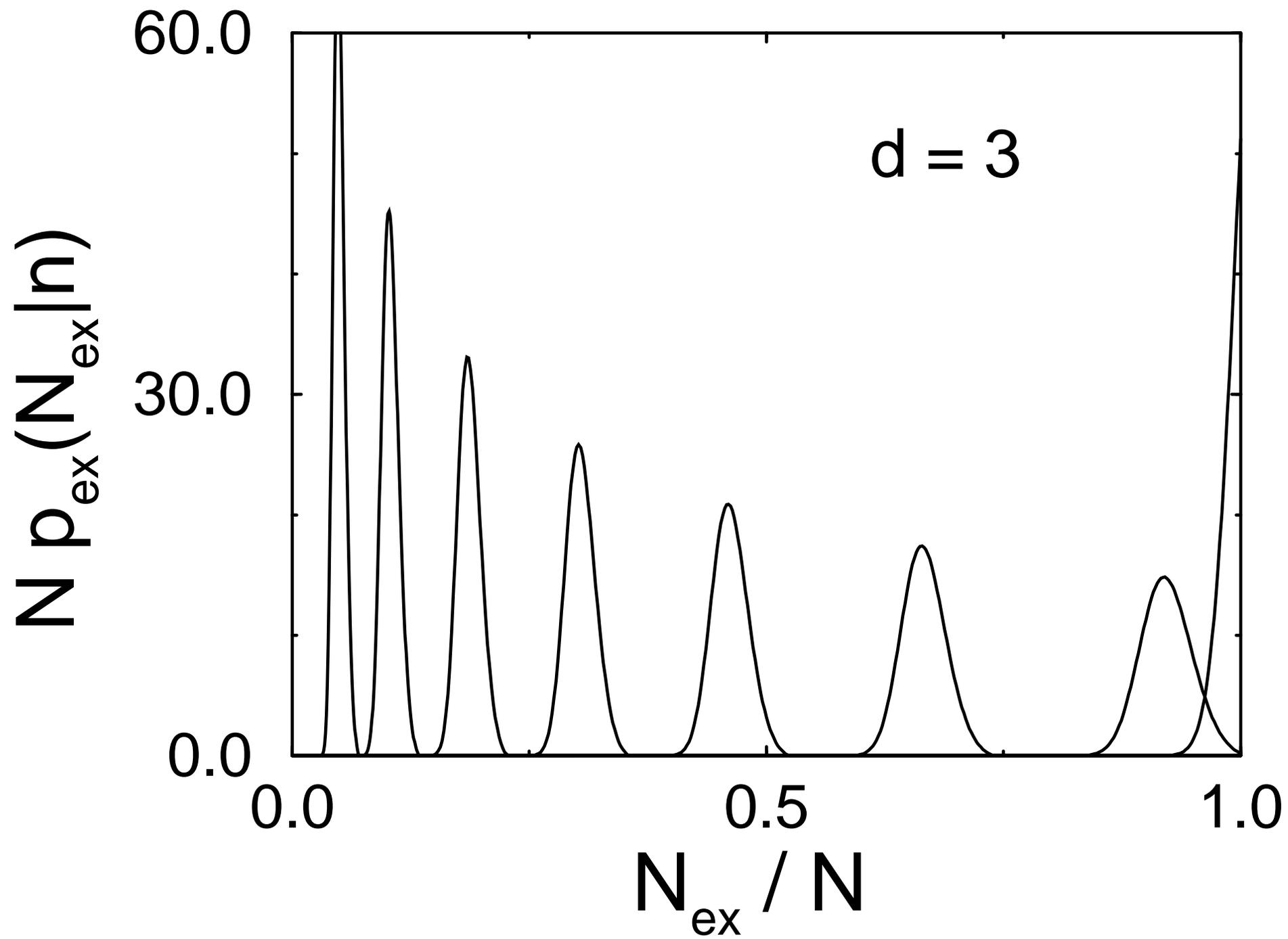
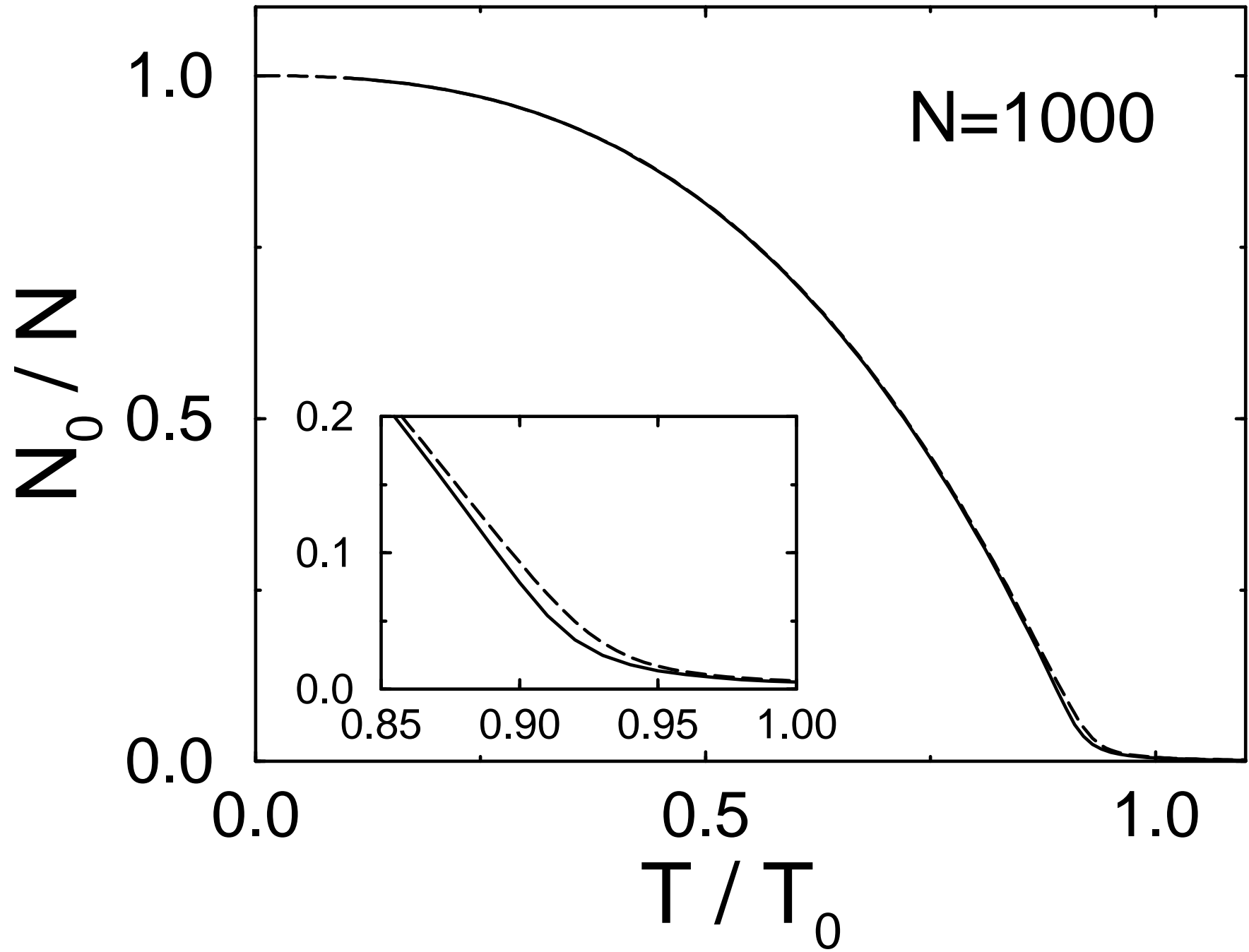
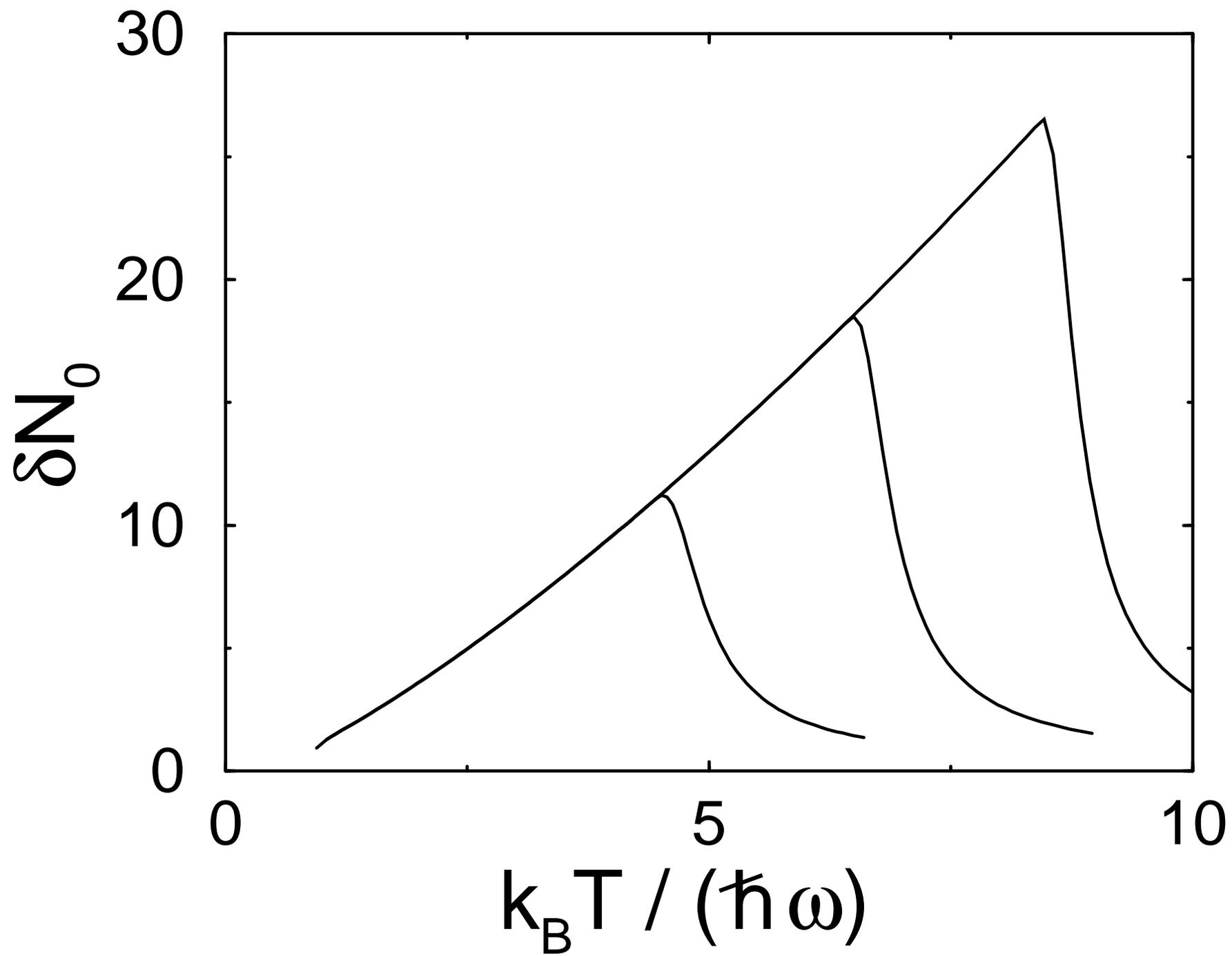
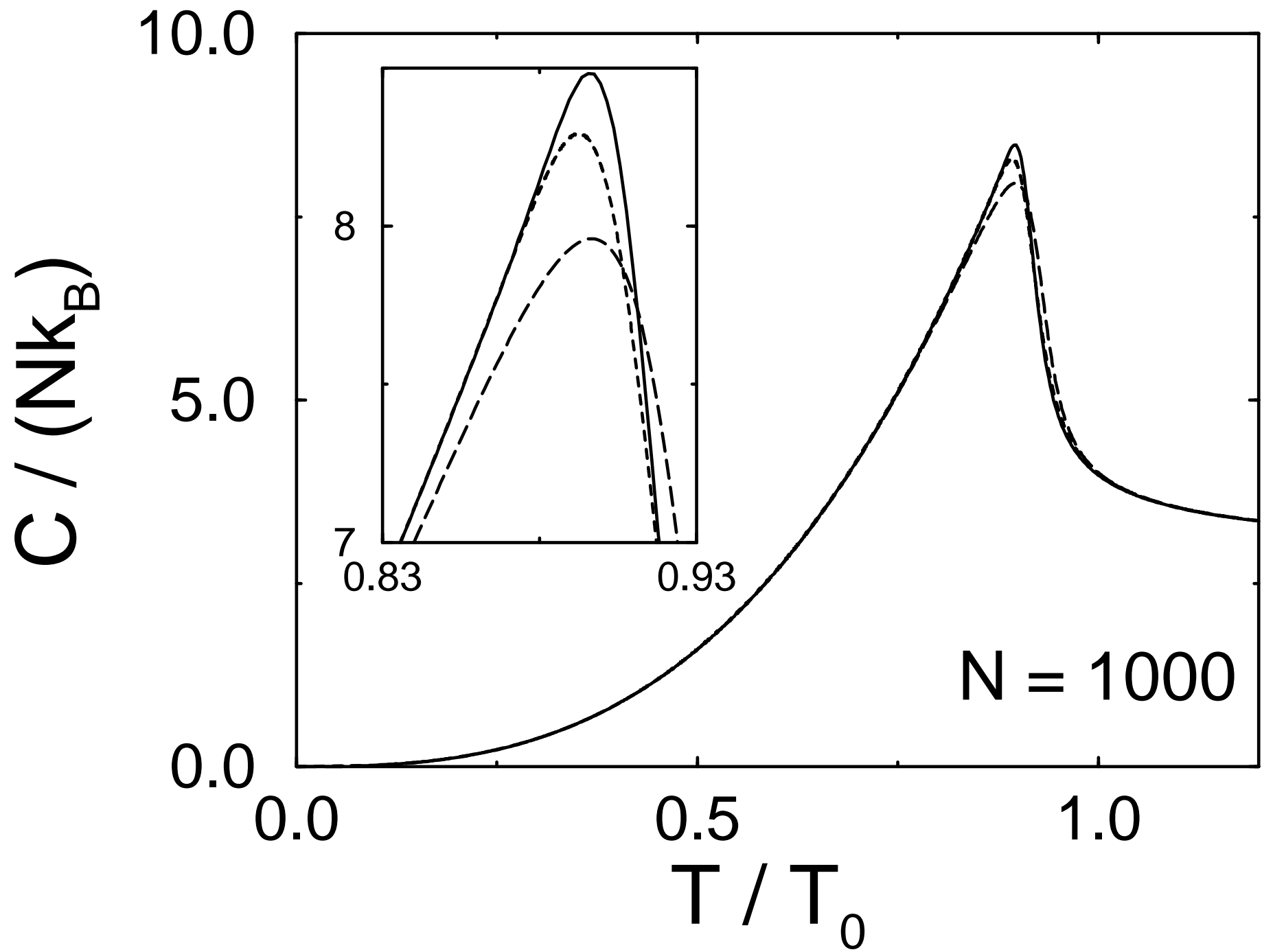
\end{document}